\documentclass[aps,twocolumn,superscriptaddress,floatfix]{revtex4-1}

\usepackage{lipsum}
\usepackage{graphicx}
\usepackage{amsmath,amssymb}
\usepackage{braket}
\usepackage{mathtools}
\usepackage{hyperref}

\usepackage{color}
\usepackage[normalem]{ulem}

\newcommand{\Z}{\mathbb{Z}}
\newcommand{\C}{\mathbb{C}}
\newcommand{\n}[1]{\left| #1 \right|}

\newcommand{\sgn}{\operatorname{sgn}}
\renewcommand{\v}[1]{\boldsymbol{#1}}
\DeclareMathOperator{\Tr}{Tr}

\usepackage{tikz}
\usetikzlibrary{shapes,calc}
\usetikzlibrary{shapes.geometric,arrows.meta,decorations.markings}

\tikzset{
	dot/.style={draw,circle,inner sep=1.5pt,fill=black},
	empty dot/.style={draw,circle,inner sep=1.5pt,fill=white},
	mid arrow/.style={postaction={decorate,decoration={
        markings,
        mark=at position .6 with {\arrow[#1,scale=1.5]{latex}}
    }}},
	spinA/.style={draw=black,thick,circle,inner sep=2.5pt, fill=figBlue},
	spinBC/.style={draw=black,thick,circle,inner sep=2.5pt, fill=figRed},
	faded/.style={opacity=0.2},
}

\definecolor{figBlue}{RGB}{32,121,180}
\definecolor{figRed}{RGB}{227,27,28}

\definecolor{darkBlue}{RGB}{140,134,248}

\pgfmathsetmacro{\s}{1.333}
\pgfmathsetmacro{\t}{0.8}
\pgfmathsetmacro{\r}{0.8}

\pgfmathsetmacro{\DY}{0.8660}
\pgfmathsetmacro{\DX}{1}
\pgfmathsetmacro{\DZ}{0.8}

\pgfmathsetmacro{\top}{0.3}
\pgfmathsetmacro{\bot}{0.3}

\begin{document}

\title{Topological Luttinger Liquids from Decorated Domain Walls}

\author{Daniel E. Parker}
\email[]{daniel\_parker@berkeley.edu}
\affiliation{Department of Physics, University of California, Berkeley, CA 94720, USA}
\author{Thomas Scaffidi}
\email[]{thomas.scaffidi@berkeley.edu}
\affiliation{Department of Physics, University of California, Berkeley, CA 94720, USA}
\author{Romain Vasseur}
\email[]{rvasseur@umass.edu}
\affiliation{Department of Physics, University of Massachusetts, Amherst, MA 01003, USA}

\date{\today}

\begin{abstract}

	We introduce a systematic construction of a gapless symmetry protected topological phase in one dimension by ``decorating'' the domain walls of Luttinger liquids. The resulting strongly interacting phases provide a concrete example of a gapless symmetry protected topological (gSPT) phase with robust symmetry-protected edge modes. Using boundary conformal field theory arguments, we show that while the bulks of such gSPT phases are identical to conventional Luttinger liquids, their boundary critical behavior is controlled by a different, strongly-coupled renormalization group fixed point. Our results are checked against extensive density matrix renormalization group calculations.

\end{abstract}
\maketitle

\section{Introduction}

Topological materials offer a rich and diverse set of condensed matter systems, extending beyond the Landau paradigm of symmetry-breaking order. The discovery of topological insulators and superconductors in the past decade~\cite{PhysRevLett.95.226801,Bernevig1757,Konig766,PhysRevLett.98.106803,PhysRevB.75.121306,PhysRevB.79.195322,Hsieh:2008aa,Hasan2010,Rasche2013,story2012} led to the general concept of Symmetry-Protected Topological phases (SPTs) which, as the name suggests, are gapped quantum phases of matter with topological properties protected by symmetry~\cite{PhysRevB.80.155131,PhysRevB.84.235128,PhysRevB.83.075102,PhysRevB.83.075103,Chen2011b,Pollmann2012,YuanMing2012,Levin2012,Chen1604,Chen2011}. They feature short-range entanglement and spectral gaps, and have gapless edge states protected by the presence of certain symmetries. (For a review, see~\cite{doi:10.1146/annurev-conmatphys-031214-014740}.) Examples include the experimentally accessible Haldane phase in quantum spin chains~\cite{PhysRevLett.61.1029,Kennedy1992,Buyers1986}. Non-perturbative techniques have led to an essentially exhaustive classification of gapped bosonic~\cite{Chen1604,Chen2011,YuanMing2012,PhysRevB.91.134404} and, to some extent, fermionic SPT phases~\cite{PhysRevB.81.134509,PhysRevB.83.075103,PhysRevB.89.201113,Wang629,PhysRevB.90.115141,Gu2015}. A key feature in the definition of SPTs is the spectral gap to bulk excitations, whose presence is generally believed to be necessary to ensure protection of the edge states.

Attention has recently turned to gapless topological materials, including Weyl and Dirac semi-metals with topologically-protected Fermi arc surface states~\cite{PhysRevB.83.205101,PhysRevX.5.031013,Lu622,Xuaaa9297,doi:10.1146/annurev-conmatphys-031113-133841}.
In previous work~\cite{scaffidi2017gapless}, the authors showed that the assumption of a spectral gap for strongly interacting SPT systems is unnecessary (see also~\cite{2012arXiv1206.1332G,PhysRevLett.118.087201}). We provided a general construction of systems ---in any dimension--- that are gapless in the bulk with symmetry-protected topological edge modes, which we dub gapless SPTs (gSPTs). Through exactly solvable examples, we demonstrated that many of the tools regularly applied to gapped SPTs --- including unitary twists \cite{Chen2011}, entanglement spectra \cite{PhysRevLett.101.010504}, strange correlators~\cite{Cenke2014}, and trial wavefunctions \cite{PhysRevB.93.115105} --- carry over to the gapless case. In general, our construction works by ``twisting''~\cite{Chen2011} a gapless topologically trivial system, or by ``decorating'' its domain walls~\cite{AshvinDecorated}. The resultant non-trivial gSPT is as stable as the underlying gapless system. Therefore twisting can turn critical points or lines into points or lines of gSPTs, but also gapless phases into gSPT phases.

In this work we systematically study an exactly solvable gSPT phase created by twisting a Luttinger liquid~\cite{0022-3719-14-19-010,giamarchi2003quantum} (LL). The resulting phase, which we call ``LL$^{\star}$'' is a topologically non-trivial phase with gapless edge modes. Just as for a regular SPT, the edge states are robust to all perturbations that preserve certain protecting symmetries. Additionally, we show that while the LL and LL$^{\star}$ phases are identical in the bulk, the boundary critical behavior of the LL$^{\star}$ state realizes a different strongly-coupled renormalization group fixed point. The spectrum of boundary critical exponents is exactly described within boundary CFT~\cite{cardy_conformal_1984,cardy_boundary_1989}, and the topological edge states lead to an exotic ``superposition'' of conformally invariant boundary conditions.

The goal of this work is two-fold. First, we wish to put the notion of topological Luttinger liquids in a more systematic context. Similar kinds of topological Luttinger liquids have been studied previously in several works~\cite{Tsui2015330,PhysRevB.91.235309,ruhman2017topological} with a variety of setups --- relying essentially on the spin-charge separation property of Luttinger liquids. 
By examining a topological Luttinger liquid as a gSPT, we can construct such gapless topological states in a more systematic way, and we can understand their topological nature using the powerful tools and techniques of SPTs. Second, by studying this one-dimensional example --- where exact results are available through CFT and explicit numerical confirmation is possible --- we may completely understand this example of a gSPT. We expect that many of the results and physical pictures developed here have close analogues in higher-dimensional gSPTs.  

This paper is organized as follows. In Section II we introduce our model and briefly describe its topological and spectral properties. Section III establishes the existence of a gSPT phase and analyzes its phase diagram. The surface critical behavior of this problem is analyzed in terms of BCFT in Section IV, along with extensive numerical validation. We then conclude with possible generalizations to higher dimensions and implications for gapless topological systems.

\section{Model and topological edge modes}
\label{Sec:Model}

\begin{figure*}
	\center
	\includegraphics{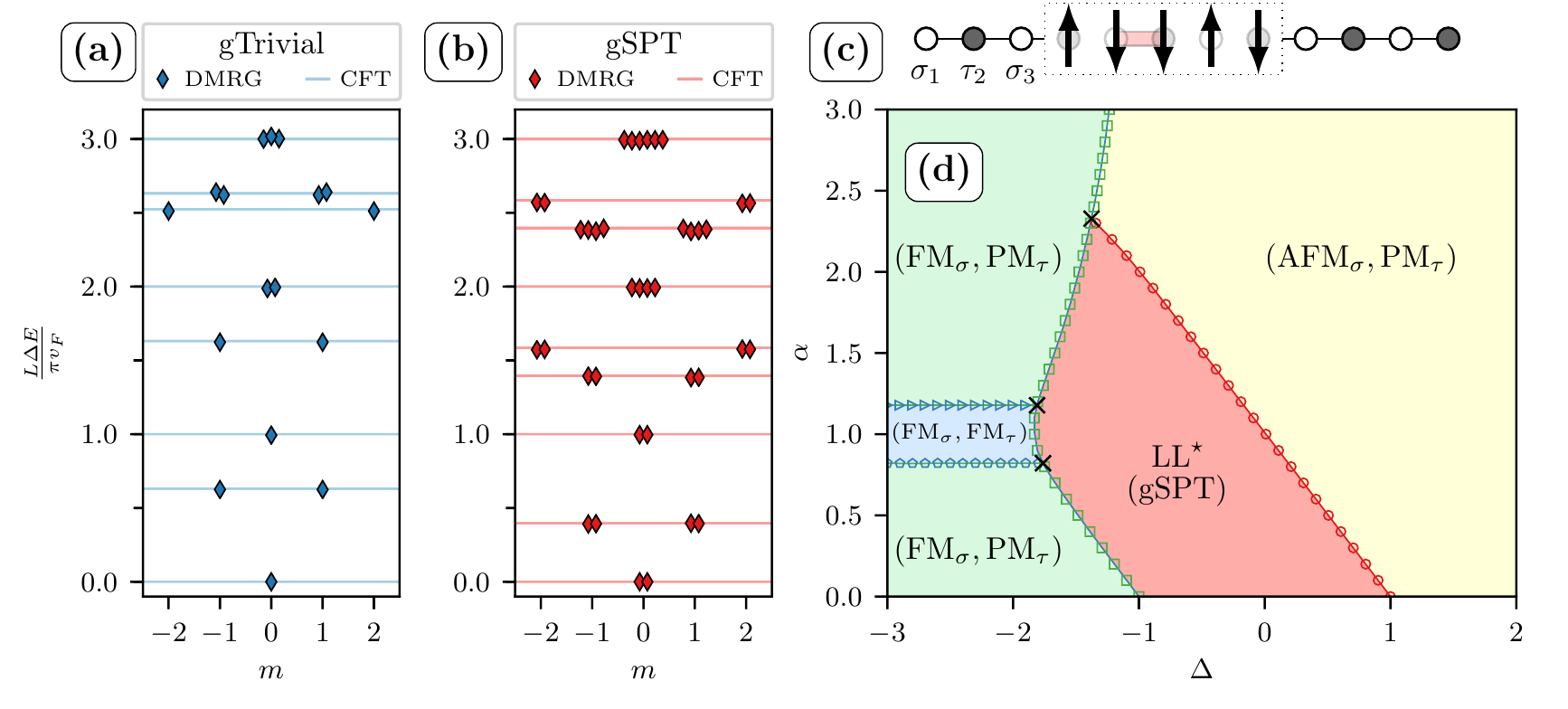}

	\caption{(a) The spectrum of $H_{\text{gTrivial}}$. (b) The spectrum of $H_{\text{gSPT}}$ for the $(Y)$-model. For both cases, open boundary conditions were chosen, and the spectra are normalized to be able to read off CFT operator dimensions. The conformal blocks are labelled by the magnetic charge sector $m$ and spaced horizontally, and small horizontal spacings show degenerate eigenvalues (up to exponential splitting). One can see that the states in the gSPT case are all doubly degenerate, due to the edge modes, but also that operator dimensions have changed relative to the gTrivial case. The numerical spectra were computed via DMRG \cite{ITensor} on up to 32 sites with finite-size scaling, and the solid lines correspond to the exponents expected from boundary CFT using $\Delta_{\rm eff} = - \cos \pi g$. To improve convergence, the gap on the paramagnetic sector was  increased from one to ten. See Section \ref{sec:BCFT} for additional numerical details. The Hamiltonian parameters used are $\Delta = 0.3, \alpha = 0.1$, $g_\tau = 0.3$, $u_\tau = 0.1$ for both (a) and (b). (c) The spin chain for the gSPT. White sites corresponds to $\sigma$ spins and grey sites to $\tau$ spins. The dotted box shows the effect of the local unitary $U$, which gives a phase factor of $-1$ for each adjacent pair of down spins (red bar). (d) The bulk phase diagram of $H_{\text{gSPT}}^{\text{LL}^{\star}}$, as computed via DMRG \cite{ITensor}. Each line denotes a different eigenvalue crossing which accompanies a phase transition, and black crosses denote multicritical points. The parameters are given by $g_\tau = 0.3$ and $u_\tau = 0.1$.
}
	\label{fig:phase_diagram}
\end{figure*}

Consider a spin-$1/2$ chain with two alternating species, which we call $\sigma$ and $\tau$, as shown in Figure \ref{fig:phase_diagram} (c). We will follow the systematic procedure for constructing a gSPT from our previous work~\cite{scaffidi2017gapless}: start with a gapless phase and then twist it by a local unitary operator to arrive at a topologically non-trivial phase. This is directly analogous to how the $\Z_2 \times \Z_2$ gapped SPT~\cite{Chen2011,AshvinDecorated,Bahri:2015aa,PhysRevB.93.155131}, the simplest one-dimensional bosonic SPT, is constructed in the decorated domain wall picture~\cite{AshvinDecorated}.

We start from a topologically trivial phase which is the XXZ model for the $\sigma$ spins and a paramagnet for the $\tau$ spins:
\begin{equation}
	\begin{aligned}
	H^{(Z)}_\text{gTrivial} \ &=\ \sum_i \sigma_i^x \sigma_{i+2}^x + \sigma_i^y \sigma_{i+2}^y + \Delta \sigma_i^z \sigma_{i+2}^z  - \tau^x_i + H_\text{int}\\
	H_\text{int} \ &=\  \sum_i \alpha \sigma_i^z \tau_{i+1}^x \sigma_{i+2}^z - 
g_\tau \tau^z_i \tau^z_{i+2} - u_\tau \tau^x_{i} \tau^x_{i+2}.\\ 
\end{aligned}
	\label{eq:H_gTrivial}
\end{equation}
This has a $U(1) \rtimes \Z_2^{(\sigma)} \times \Z_2^{(\tau)}$ global symmetry, generated by $U_\theta = \prod_{i=1,3,\dots} e^{i \theta \sigma^z_i/2}$, $\mathcal{C}_\sigma = \prod_{i=1,3,\dots} \sigma_i^x$ and $\mathcal{C}_\tau = \prod_{i=2,4,\dots} \tau_i^x$ respectively.

The choice of anisotropy in the $Z$-direction is arbitrary; taking the anisotropy in the $Y$-direction gives the same gTrivial state, but we will show below that it will produce a slightly different gSPT. Because this choice amounts to an onsite unitary $\sigma_i^y \leftrightarrow \sigma_i^z$, we will, for the most part, treat both models simultaneously. For brevity, we refer to the Hamiltonian in Eq.~\eqref{eq:H_gTrivial} as the $(Z)$-model, and the one obtained from Eq.~\eqref{eq:H_gTrivial} by doing $\sigma_i^y \leftrightarrow \sigma_i^z$ as the $(Y)$-model:
\begin{equation}
	\begin{aligned}
	H^{(Y)}_\text{gTrivial} \ &=\ \sum_i \sigma_i^x \sigma_{i+2}^x + \sigma_i^z \sigma_{i+2}^z + \Delta \sigma_i^y \sigma_{i+2}^y  - \tau^x_i + H_\text{int}\\
	H_\text{int} \ &=\  \sum_i \alpha \sigma_i^y \tau_{i+1}^x \sigma_{i+2}^y - 
g_\tau \tau^z_i \tau^z_{i+2} - u_\tau \tau^x_{i} \tau^x_{i+2}.\\ 
\end{aligned}
	\label{eq:H_gTrivialY}
\end{equation}
 However, the slight differences between the two models will help to elucidate several subtle issues, so we will contrast them at several points below.

 The symmetry protecting the topological properties is given by $\Z_2^{(\sigma)} \times \Z_2^{(\tau)}$, while the $U(1)$ symmetry enforces a gapless bulk in the Luttinger liquid phase. Interactions in $H_\text{int}$ are chosen to be the simplest ones compatible with the symmetry. The perturbation $g_\tau$ takes the $\tau$'s from a simple paramagnet to an Ising model, and $u_\tau$ takes the Ising model away from integrability. The interaction $\alpha$ couples the $\sigma$ and $\tau$ sectors together by the simplest term compatible with all symmetries.

To twist the model into a non-trivial gapless SPT, we use the local unitary operator $U$ (also used to construct the gapped $\Z_2 \times \Z_2$ SPT) 
\begin{equation}
U = \prod_{{\rm DW}(\sigma)} (-1)^{\frac{1-\tau_i^z}{2}},
\label{eq:twist_operator}
\end{equation}
where the product runs over all the domain walls of the $\sigma$ spins in the $z$ basis. One can think of $U$ as attaching a charge of $\Z_2^{(\tau)}$ to the domain walls of $\Z_2^{(\sigma)}$~\cite{AshvinDecorated,PhysRevB.91.195117}. From a more concrete perspective, $U$ gives a factor of $(-1)$ for each instance of two consecutive down spins (in the $z$ basis) in a classical spin configuration, as shown in Figure~\ref{fig:phase_diagram}~(c).

The non-trivial gapless SPT is defined by 
\begin{equation}
	H_{\text{gSPT}} = U H_\text{gTrivial} U.
	\label{eq:H_gSPT}
\end{equation}
For the $(Z)$-model, the $U(1)$ symmetry is invariant under conjugation by $U$, but for the $Y$-model, the $U(1)$ symmetry is also twisted in the process, and is therefore slightly less natural as a physical symmetry after twisting.  
With periodic boundary conditions, $H_\text{gSPT}$ clearly has the same spectrum as $H_\text{gTrivial}$. However, for open boundary conditions, we will see that $H_\text{gSPT}$ has gapless edge modes. 
This manifests as a two-fold exponential degeneracy of all the low-energy states, as shown in Figure \ref{fig:phase_diagram} (b). Remarkably, we will see that these topological edge modes (and the associated two-fold degeneracy of the spectrum) are robust to symmetry-preserving perturbations, even if they are strongly coupled to the bulk gapless degrees of freedom.

To demonstrate the presence of edge modes, let us first restrict ourselves to a limit were they can be found exactly: a semi-infinite chain $i\ge 1$, starting with $\sigma_1$, without interactions, i.e. $\alpha = g_\tau = u_\tau = 0$. The action of $U$ on the Pauli matrices is $\sigma_i^x \to \tau_{i-1}^z \sigma_i^x \tau_{i+1}^z, \sigma_i^y \to \tau_{i-1}^z \sigma_i^y \tau_{i+1}^z, \sigma_i^z \to \sigma_i^z$, and similarly for the $\tau$'s. Hence, the first few terms in $H_\text{gSPT}$ for the $(Z)$-model are given by 
\begin{equation}
	H_\text{gSPT}^{\text{open}} =
		\Delta\sigma_1^z \sigma_3^z 
		+ \tau_2^z \sigma_3^x \sigma_5^x \tau_6^z
        + \tau_2^z \sigma_3^y \sigma_5^y \tau_6^z
        - \sigma_1^z \tau_2^x \sigma_3^z
        +\cdots
\end{equation}
(Terms such as $\sigma_1^x \sigma_3^x\tau_4^z$ and $\sigma_1^y \sigma_3^y\tau_4^z$ are not compatible with the symmetries and cannot be included.) In this simple case, the edge mode $\Psi$ is simply the first spin; $\Psi = \sigma_1^z$ commutes with the Hamiltonian and thus indexes a double degeneracy of every state. Even in this simple, exactly solvable limit, the edge mode does {\em not} completely decouple from the gapless degrees of freedom. In particular, because of the $\Delta \sigma_1^z \sigma_3^z $ term in the Hamiltonian, enforcing $\sigma_1^z = \pm 1$ induces an effective magnetic field $\pm \Delta$ acting on $\sigma_3^z$. The effects of this boundary magnetic field on the gapless degrees of freedom will be discussed in Section~\ref{sec:BCFT}. We emphasize that this is dramatically different from edge modes at the boundary of {\em gapped} SPTs: for gapped SPTs, one can construct local operators that can flip the edge mode from $\sigma_1^z = \pm 1$ to $\sigma_1^z = \mp 1$, while in the gapless case above this is not possible because of the coupling to the gapless bulk modes. (In particular, this is why there is only a two-fold degeneracy of the low-energy states for an open system of finite length $L$, instead of the four-fold degeneracy expected for gapped 1d SPT with two spin-$\frac{1}{2}$ edge modes.)

	The edge modes survive in the presence of generic interactions compatible with the symmetry. In particular, we add $\alpha, g_\tau$, and $u_\tau$, as well as additional $\Z_2^{(\sigma)} \times \Z_2^{(\tau)}$-preserving  boundary perturbations such as $\sigma_1^x$. (We allow terms breaking the $U(1)$ symmetry at the boundary since this symmetry only protects the gaplessness of the bulk, and plays no role in the topological properties of the system.) Due to the spectral gap in the $\tau$ sector, local channels to flip the edge mode are exponentially suppressed. Thus the only effect of interactions is to dress $\Psi$, which remains largely localized to the first site. In a system with finite length $L$, open boundary conditions and say, $u_\tau=\alpha=0$ but  $g_\tau  \ll 1$, the amplitude for flipping the edge modes at both ends scales as ${\sim} g_\tau^L = {\rm e}^{L \log g_\tau }$ and is thus exponentially small in $L$. As for gapped SPTs, we thus expect the edge modes to remain protected by the gap of the $\tau$ spins away from the exactly solvable limit described above. This picture is confirmed numerically below.

We may thus consider $H_\text{gSPT}$ as a gapless symmetry-protected topological state. We emphasize that our construction was completely systematic, following the general construction from previous work~\cite{scaffidi2017gapless} and employing the same unitary transformation used for the standard $\Z_2 \times \Z_2$ SPT. We should therefore expect it to inherit many of the properties of normal SPTs. As a practical matter, this implies $H_\text{gSPT}$ should be robust to perturbations and therefore constitute a phase. We confirm this in the next section.

\section{Phase diagram}

In this section we study the phase diagram of $H_\text{gSPT}$, and show that it has a genuine gapless SPT \emph{phase} protected by the symmetry $U(1) \rtimes \Z_2^{(\sigma)} \times \Z_2^{(\tau)}$. Since, for periodic boundary conditions, $H_\text{gSPT}$ and $H_\text{gTrivial}$ are related by a local unitary transformation, they share the same bulk properties. In particular, their phase boundaries are the same; the LL$^{\star}$ state is as stable as the original (untwisted) LL phase~\cite{scaffidi2017gapless}. We therefore analyze Eq.~\eqref{eq:H_gTrivial} to understand the phase diagram of $H_\text{gSPT}$ (which is the same for both the $(Y)$- and $(Z)$-models).

	A numerical phase diagram is shown in Figure \ref{fig:phase_diagram} (d), computed with DMRG~\cite{white92,ITensor} and finite-size scaling. To avoid fine-tuning, we include all perturbations from $H_\text{int}$. It is clear that LL$^\star$ is a genuine phase and not just a critical point. By working in various limits and approximations, we may identify the various proximate phases and many of the phase transitions. First, the line $\alpha = 0$ is the same as the well-known XXZ model, which is a ferromagnet for $\Delta < -1$ and an antiferromagnet for $\Delta > 1$. The middle region $-1 < \Delta < 1$ is a Luttinger liquid, whose low-energy properties are given by the compact free boson conformal field theory, a fact we employ in the following section.

When $\alpha \to \infty$, the Hamiltonian is dominated by the interaction $\alpha \sigma^z \tau^x \sigma^z$. This has $2^{L/2}$ degenerate ground states: any classical configuration of $\sigma^z$ is permitted, and the $\sigma$ spins fix the $\tau$ spins by the requirement  $\braket{\sigma^z_i \tau^x_{i+1} \sigma^z_{i+2}} = -1$. 
	Adding the other terms at first order in perturbation theory gives an effective Hamiltonian for this low-energy sector of $H_\text{eff} = \sum_i \left( \Delta +1 \right) \sigma^z_i \sigma_{i+2}^z$. This implies the ground state is ferromagnetic for $\Delta < -1$ and undergoes a direct transition to the anti-ferromagnetic phase at $\Delta > -1$. Indeed, we see a transition near $\Delta = -1$ for $\alpha \gg 0$ in the phase diagram.

For $\Delta \to \infty$, the situation is trivial and we always have an anti-ferromagnet for $\sigma$ and a paramagnet for the $\tau$'s. In the opposite limit $\Delta \to - \infty$, we may assume the $\sigma$'s are perfectly ferromagnetic, and hence $\braket{\sigma^z_i \sigma^z_{i+2}} = 1$. This gives an effective Hamiltonian $H_\text{eff} = -\sum_i (1-\alpha) \tau^x_{i} + g_\tau \tau^z_i \tau^z_{i+2} + u_\tau \tau^x_i \tau^x_{i+2}$ for the $\tau$ spins. When $u_\tau = 0$, this gives Ising transitions for the $\tau$'s at $\alpha = 1 \pm g_\tau$. Adding in $u_\tau$ breaks the integrability of the Ising model, moving the transitions to $\alpha = 1 \pm \left( g_\tau - O(u_\tau) \right)$.

Lastly, we can understand the phase transitions out of the LL$^{\star}$ phase in an effective theory. Integrating out the gapped $\tau$ spins gives
$\Delta_\text{eff}(\alpha) = \Delta  + \alpha \braket{\tau^x} \approx\Delta + \alpha$. This gives phase transitions at $\Delta_\text{eff}(\alpha) = \pm 1$, i.e. $\Delta(\alpha) = \pm 1 - \alpha$. Figure \ref{fig:phase_diagram} (d) shows this argument accurately predicts the phase boundaries for LL$^{\star}$ up to $\alpha \approx 1$.

We have established that LL$^\star$ is a genuine phase within our choice of parameters and does not rely on fine-tuning. The stability of LL$^\star$ is a separate question, which we defer to Section \ref{sec:BCFT_LLstar}. For now, note that both the LL and LL$^\star$ phases can be gapped out by dimerizing the $\sigma$ spins, so translation invariance by a single unit cell should also be included in the symmetry group protecting the gaplessness of the system.

\begin{figure*}
	\centering
	\includegraphics{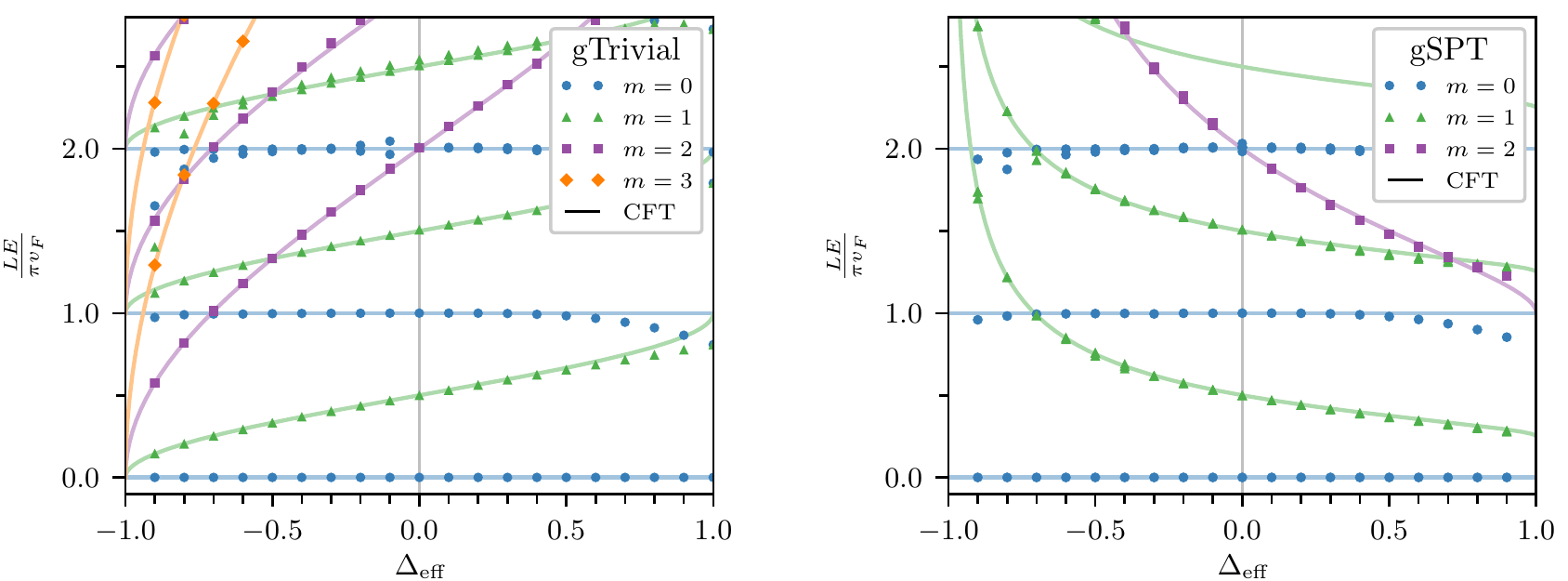}
	\caption{(Left) Comparison between the BCFT predictions from \eqref{eq:gTrivial_spectrum} for the trivial XXZ spectrum with open boundary conditions (solid lines) and numerical eigenvalues from DMRG (data). (Right) Comparison between BCFT predictions from \eqref{eq:gSPT_spectrum}, and DMRG for $H_\text{gSPT}^{(Y)}$. Numerical details are in the text. The degeneracies of the different branches can be inferred from Fig.~\ref{fig:phase_diagram}.
}
	\label{fig:spectrum_comparison}
\end{figure*}

\section{Boundary criticality}
\label{sec:BCFT}

Since $H_\text{gSPT}$ is a gapless model in 1+1 dimensions, its universal properties are described by conformal field theory~\cite{DiFran}. In this section we will show that boundary CFT (BCFT)~\cite{cardy_conformal_1984,cardy_boundary_1989} predicts the exact low-energy spectrum (below the paramagnetic gap) in the LL$^\star$ phase, including degeneracies. We will see that $H_\text{gSPT}$ realizes a superposition of boundary conditions that would otherwise require symmetry-breaking fields in non-topological (gTrivial) systems.   

\subsection{BCFT for LL} 

To start, let us review some facts about the standard (non-twisted) XXZ model, which will apply to $H_\text{gTrivial}$. Through bosonization, and integrating out the $\tau$ spins, the low-energy physics of this model can be captured by a single gapless bosonic field (Luttinger liquid)
\begin{equation}
	{\cal L}_{\rm LL} = \frac{g}{4 \pi} \left( \partial_\mu \varphi  \right)^2,
	\label{eq:XXZ_bosonized}
\end{equation}
where $\varphi$ is a free scalar field compactified onto a circle of radius $2\pi$ (i.e. $\varphi \equiv \varphi + 2\pi$), and $g = \pi^{-1} \arccos(-\Delta_{\rm eff})$ is the Luttinger parameter~\cite{giamarchi2003quantum} and $\Delta_{\rm eff}$ is the effective anisotropy of the $\sigma$ spins after integrating out the gapped $\tau$ spins.  (For our choice of parameters, $\Delta_\text{eff} \approx \Delta +\alpha \braket{\tau^x} $.) For simplicity, we rescaled the space coordinate $x$ to set the Fermi velocity $v_F = 2\pi \frac{\sqrt{1-\Delta_{\rm eff}^2}}{\arccos \Delta_{\rm eff}}$ to unity in the Lagrangian. 
 The correspondence between spins and the bosonization fields is~\cite{giamarchi2003quantum}
\begin{align}
	\sigma_k^z &\simeq \frac{1}{ \pi} \partial_x \varphi+ A (-1)^k \sin \varphi, \notag\\
	\sigma_k^x \pm i \sigma_k^y &\simeq e^{\pm i \theta} \left( B \cos \varphi + C (-1)^k  \right),
	\label{eq:spin_field_correspondence}
\end{align}
where $A,B,C$ are non-universal constants, and $\theta$ is the dual variable with commutation relation $[\varphi(x),\theta(y)] = i \pi \sgn(x-y)$. Free boundary conditions on the $\sigma$ spins correspond to the Dirichlet boundary conditions $\varphi(x=0) = 0 = \varphi(x=L)$~\cite{oshikawa1997boundary}. The $U(1) \rtimes \Z_2$ symmetry of the $\sigma$ spins acts as $\theta \to \theta + \alpha$, and $(\varphi,\theta) \to (-\varphi,-\theta)$, respectively. The gaplessness of the XXZ chain is also protected by translation invariance, which acts on the field theory~\cite{PhysRevB.46.10866,giamarchi2003quantum} as a discrete symmetry $(\varphi,\theta) \to (\varphi+\pi,\theta+\pi)$ (or by a mirror symmetry $(\varphi,\theta) \to (\pi-\varphi,\theta+ \varphi)$). With these protecting symmetries, the leading perturbation is $\cos 2 \varphi$ with scaling dimension $2/g$, which is irrelevant for $\Delta_{\rm eff}<1$ and opens a N\'eel gap for $\Delta_{\rm eff}>1$. The same stability analysis applies to the LL$^\star$ (gSPT) state.

The partition function of this gTrivial model can be computed using standard tools of boundary conformal field theory and is summarized in the Appendix. Ignoring non-universal contributions, it reads
\begin{equation}
	Z_\text{gTrivial}(g) = \Tr \ {\rm e}^{-\beta H_{\rm gTrivial}} = \frac{1}{\eta(q)} \sum_{e \in \Z} q^{g e^2}, 
	\label{eq:Z_gTrivial}
\end{equation}
where $q = e^{- \beta\pi v_F/L}$ and $\eta(q) = q^{1/24} \prod_{n=1}^\infty (1-q^n)$ is the Dedekind $\eta$-function. This partition function can be interpreted as a generating function $q^{-1/24} \sum_i q^{\Delta_i}$ for the boundary scaling dimensions $\Delta_{e,n} = g e^2 + n$ ($e\in \Z, n>1$)  of the theory, whose multiplicity is $p(n)$, the number of integer partitions of $n$~\cite{DiFran}. In CFT language, these are the scaling dimensions of the electric operators for the massless free boson with Dirichlet boundary conditions (and their descendants). The value of $e$ for a state corresponds to total spin quantum number $\braket{\sigma^z} = \sum_i \braket{\sigma^z_i}$. These critical exponents can also be extracted numerically from the finite size scaling of the energy gaps of $H_{\rm gTrivial}$ 
\begin{equation}
	E_{e,n}(L)-E_0(L) = \frac{\pi v_F}{L}\Big[ g(\Delta_\text{eff}) e^2 + n \Big],\text{ for } e \in \Z,\; n > 1,
	\label{eq:gTrivial_spectrum}
\end{equation}
with $E_0(L)$ is the groundstate energy and $E_{k,n}(L)$ the energy of an excited state.   To compare with numerics, the standard technique is to determine the operator dimensions via extrapolation from finite size data. In practice, one extracts the operator dimension $x$ by fitting $\frac{\pi v_F}{L} \left[ E(L) - E_0(L) \right] = x + a_1 L^{-1} + a_2 L^{-2} + \cdots$. We perform this in Figures \ref{fig:phase_diagram} (a) and \ref{fig:spectrum_comparison} (a). Figure \ref{fig:spectrum_comparison} (a) shows \eqref{eq:gTrivial_spectrum} matches the numerical data essentially exactly, with minor deviations for $\n{\Delta_{\rm eff}} \approx 1$ where additional operators become marginal.

Numerics were performed with DMRG using the ITensor Library~\cite{ITensor}. For each $\Delta_{\rm eff}$, the lowest 50 eigenvalues were computed for $L = 20,24,28,32$ with bond dimension up to $\chi = 400$ or until the error, estimated by $\braket{\psi|H^2|\psi} - \braket{\psi|H|\psi}^2$, was less than $10^{-6}$ for each state. The Hamiltonian parameters used were $\alpha = 0.1$, $g_\tau = 0.3, u_\tau = 0.1$. To improve convergence under finite size scaling, the gap to the $\tau$ sector was increased by taking $-\tau_i^x \to -10 \tau_i^x$. 
The data shown in Figure \ref{fig:phase_diagram} (a) and (b) is a slice of Figure \ref{fig:spectrum_comparison} at $\Delta = 0.3$. The degeneracies, as well as the operator dimensions, match CFT predictions. Away from $\n{\Delta} \approx 1$, errors are ${\sim} 1\%$, and are due to a combination of finite-size fitting error, failure to match corresponding eigenvalue correct at different finite sizes, and DMRG errors. The last is increasingly prevalent beyond eigenvalue 30 where errors in previous states begin to accumulate significantly. 

\subsection{BCFT for LL$^{\star}$}
\label{sec:BCFT_LLstar}

We now consider the case of the non-trivial gSPT Luttinger Liquid (LL$^\star$).	Because $H_\text{gTrivial}$ and $H_\text{gSPT}$ have identical spectra with periodic boundary conditions, the difference between the two Hamiltonians can only manifest itself at the boundaries. Hence the low-energy part of the spectrum of $H_\text{gSPT}$ should again be described by the effective field theory \eqref{eq:XXZ_bosonized} --- but now with different boundary conditions. In this section we restrict ourselves to the $(Y)$-model and treat the $(Z)$-model subsequently.

To determine the correct boundary conditions, let us consider the physical effect of the topological edge modes. From the exactly solvable limit studied in Sec.~\ref{Sec:Model}, we expect LL$^\star$ to have localized topological edge modes that strongly overlap with $\sigma^z$ at the edges. To leading order, an edge mode in the states $\ket{\uparrow}$ or $\ket{\downarrow}$ will induce an effective boundary field $\pm h_b$ for the gapless $\sigma$ spins in the $z$ (easy-plane) direction. Using the bosonization formulas~\eqref{eq:spin_field_correspondence} (with the cyclic relabeling $X \to Z$, $Y \to X$, $Z \to Y$), we thus find the effective action for the LL$^\star$ phase on an interval $[0,L]$     
\begin{align}
 S_\text{gSPT} &= \frac{g}{4 \pi} \int_0^L dx \int d\tau \left( \partial_\mu \varphi  \right)^2 \mp \lambda_L \int d\tau \cos \theta(x=0) \notag\\
 &  \mp \lambda_R \int d\tau \cos \theta(x=L),
\end{align}
where the $\mp$ signs correspond to the edge modes on the left and right of the chain being in the state $\ket{\uparrow}$ or $\ket{\downarrow}$ (in the $z$-basis), respectively. With our bosonization convention, we have $\cos \theta \to -\cos \theta$ under the $\Z_2^{(\sigma)}$ symmetry, so that the $\cos \theta$ terms would not be allowed by themselves in a $\Z_2^{(\sigma)} \times \Z_2^{(\tau)}$-symmetric   action. The coefficients $\lambda_R$ and $\lambda_L$ are non-universal functions of the boundary field $h_b$ induced by the edge modes. They are simply proportional to $h_b$ for small fields. From our microscopic model, we expect $h_b$ to be the same order as the single-particle bandwidth, so the $\lambda$ couplings are not perturbatively small. In any case, the $\cos \theta$ boundary perturbations have scaling dimension $h=g <1$ and are therefore always relevant~\cite{0305-4470-31-12-003}. At low energy, $\lambda_R$ and $\lambda_L$ flow to strong coupling so the cosines will pin down the $\theta$ field at the boundary. This induce a flow from Dirichlet ($\left. \varphi \right|_\partial = 0$) to Neumann boundary conditions ($\left. \theta \right|_\partial = 0$ or $\left. \theta\right|_\partial = \pi$ depending on the $\pm$ signs).   

The total partition function in this case is given by the superposition of these conformally invariant boundary conditions corresponding to the four configurations of the two edge modes
\begin{equation}
	Z_\text{gSPT} = Z_{0,0} + Z_{0,\pi} + Z_{\pi,0} + Z_{\pi,\pi},
	\label{eq:Z_gSPT_I}
\end{equation}
where $Z_{a,b}$ denotes the boundary partition function of a free boson with $\theta(x=0)=a$ and $\theta(x=L)=b$. Using standard boundary CFT tools (see Appendix), we find that
\begin{align}
	Z_\text{gSPT}(g) &= \frac{2}{\eta(q)} \sum_{m \in \Z} \left( q^{m^2/g} + q^{\left( m-\frac{1}{2} \right)^2/g} \right) \notag  \\
    &= 2 Z_\text{gTrivial}\left(\frac{1}{4g} \right).
	\label{eq:Z_gSPT_II}
\end{align}
where, as above, $g = \pi^{-1} \arccos(-\Delta_\text{eff})$.  This immediately implies the spectrum for $H_\text{gSPT}$ is
\begin{equation}
	E_{m,n}-E_0 = \frac{\pi v_F}{L}\Big[ \frac{m^2}{4g} + n \Big],\text{ for } m \in \Z,\; n > 1,
	\label{eq:gSPT_spectrum}
\end{equation}
now with multiplicity $2p(n)$. This is the same as \eqref{eq:gTrivial_spectrum} with $g \to 1/4g$ and the double degeneracy. In the special case $\Delta_{\rm eff} = 0$ (i.e. $g = \tfrac{1}{2}$), the spectrum of $H_\text{gSPT}$ is exactly a doubly degenerate version of $H_\text{gTrivial}$. (This non-interacting limit agrees with previous results~\cite{2012arXiv1206.1332G,Grover280,scaffidi2017gapless}.) For generic $\Delta_{\rm eff}$, however, the spectrum of boundary critical exponents will change. Employing the same procedure as for the trivial case, Figure~\ref{fig:spectrum_comparison} (b) compares~\eqref{eq:gSPT_spectrum} to numerical results. One can see that the correspondence is excellent, away from $\n{\Delta_{\rm eff}} \approx 1$. Note that realizing these boundary conditions in a topologically trivial Luttinger liquid would require introducing symmetry-breaking magnetic fields at the boundary.

Note that the groundstate of $H_\text{gSPT}$ (as well as all low-lying excited states) is doubly degenerate, in contrast with ordinary gapped SPTs with spin-$\frac{1}{2}$ edge modes that exhibit a 4-fold degeneracy, such as the Haldane chain. This is because edge modes in the configurations $\Ket{\uparrow_L \downarrow_R}$ and $\Ket{\downarrow_L \uparrow_R}$ in the $z$ basis (corresponding to $Z_{0,\pi}$ and $Z_{\pi,0}$) induce a kink of $\pi$ in $\theta(x)$. In boundary CFT language, this corresponds to the insertion of a boundary condition changing (BCC) operator. This modifies the critical exponents by a factor of $-\tfrac{1}{2}$, visible in Eq.~\eqref{eq:Z_gSPT_II}. The states with $\Ket{\uparrow_L \downarrow_R}$, $\Ket{\downarrow_L \uparrow_R}$ are hence power-law (${\sim} 1/L$) split from states with $\Ket{\uparrow_L \uparrow_R}$, $\Ket{\downarrow_L \downarrow_R}$ in any finite size system. Of course, finite-size eigenstates should preserve the symmetry $\mathcal{C}_\sigma = \prod_i \sigma^x_i$ and will therefore come in cat-state superpositions of the edge modes $\Ket{\uparrow_L \downarrow_R} \pm \Ket{\downarrow_L \uparrow_R}$ and $\Ket{\uparrow_L \uparrow_R} \pm \Ket{\downarrow_L \downarrow_R}$. At finite sizes, these cat states are not exactly degenerate but are exponential split ${\sim} {\rm e}^{-L/\xi}$ since the edge modes are exponentially localized, as the degeneracy between the states is broken at $L$th order in perturbation theory. In practice, this exponential splitting is completely negligible compared to the $1/L$ critical scaling in Eq.~\eqref{eq:gSPT_spectrum}.

The presence of cat states of edge modes has interesting consequences for the edge magnetization in the groundstate. As discussed above, in the doubly degenerate (up to exponentially small correction) ground state of $H_\text{gSPT}$, the edge modes the ends of the chain are in linear combinations of $\ket{\uparrow_L \downarrow_R} \pm  \ket{\downarrow_L \uparrow_R}$ (in the $z$-basis). As in usual symmetry breaking, any magnetic field larger than the exponential splitting between these cat states and much smaller than the CFT finite size gap ${\sim} 1/L$ is enough to pick out one of the two states breaking the symmetry at the boundary.  This creates a spontaneous edge magnetization for the $\sigma$ spins in the $z$ direction, which decays algebraically into the bulk (see Fig.~\ref{fig:magnetization} (a) and (b)). 
The $\Z_2^{(\sigma)}$ symmetry is therefore spontaneously broken at the edge.
In the language of surface criticality, this would correspond to an extraordinary transition, in which the edge breaks the symmetry before the bulk does~\cite{cardy1996scaling}. 
Note that such an extraordinary transition would be forbidden in a regular 1D system since the edge is zero-dimensional.
The boundary fixed point $Z_{\rm gSPT}$ is therefore extremely unusual and strongly relies on SPT physics.

\subsection{BCFT for the $(Z)$-model}
\label{sec:whyY}

\begin{figure}
	\centering
	\includegraphics{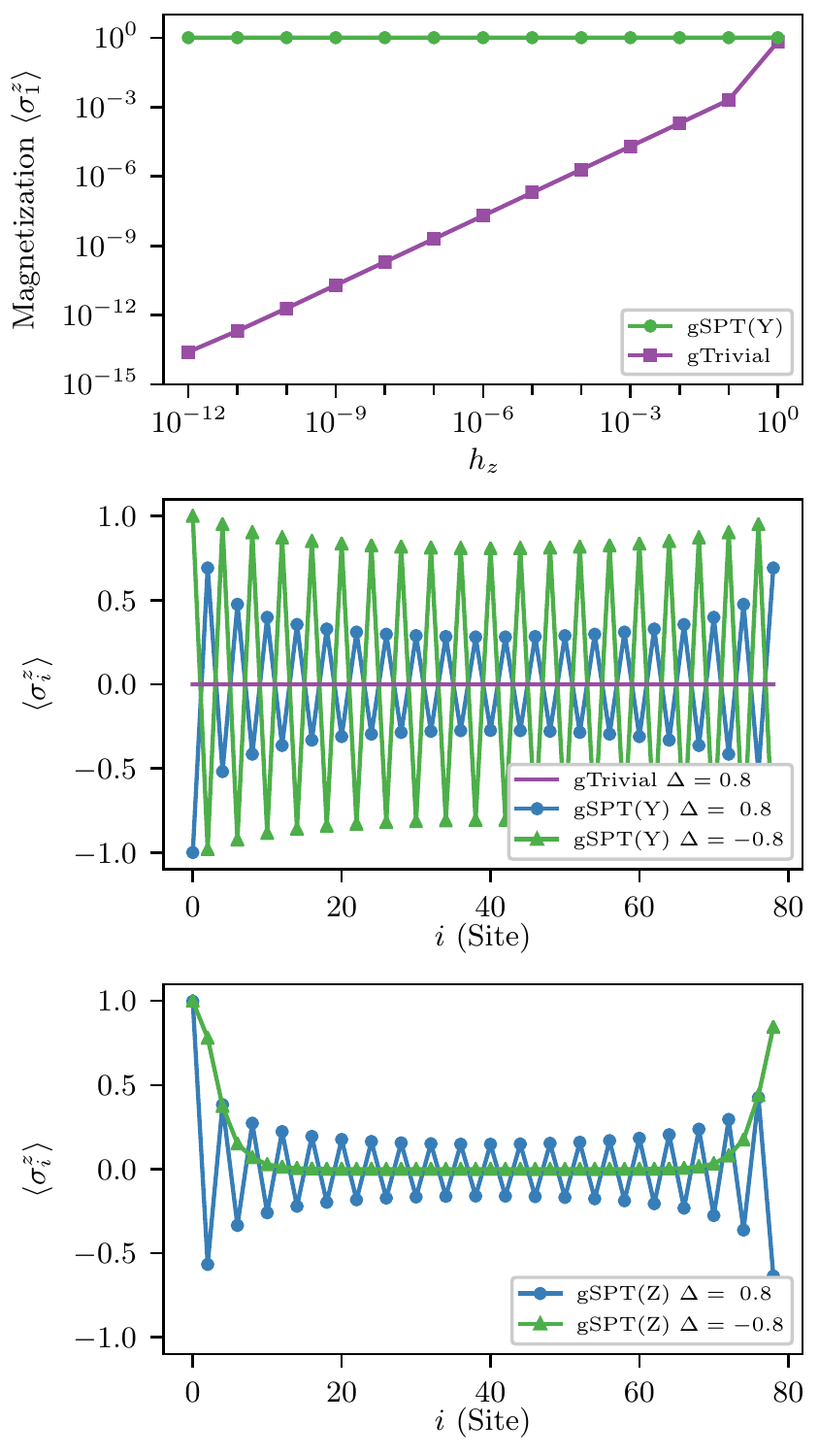}
	\caption{(Top) Edge magnetization for gSPT and gTrivial systems at $\Delta = \alpha = 0, g_\tau = 0.3,u_\tau = 0.1$ on 16 sites, via exact diagonalization. (Middle) Power-law decay of edge magnetization into the (antiferromagnetic) bulk via DMRG on 80 sites under an applied magnetic field $h_z = 10^{-5}$. The power law depends on $\Delta$; $g_\tau = 0.3, u_\tau = 0.1, \alpha = 0$.  (Bottom) Edge magnetization for $H_\text{gSPT}^{(Z)}$ with the same parameters. The magnetization is now non-universal and changes between algebraic antiferromagnetism for $\Delta = 0.8$ and exponentially decaying ferromagnetism for $\Delta =-0.8$.
}
	\label{fig:magnetization}
   
\end{figure}

Now that the boundary CFT picture of the edge modes is well-established for the $(Y)$-model, let us comment briefly on the $(Z$)-model. The same techniques apply, but the boundary critical exponents happen to be non-universal in this case. The ground states of  $H_\text{gSPT}^{(\text{$Z$})}$ also have pinned edge spins along the $z$ axis, which are antiferromagnetic (i.e. $\ket{\uparrow_L \downarrow_R}$ and $\ket{\downarrow_L \uparrow_R}$) for $\Delta_{\rm eff} > 0$ and ferromagnetic (i.e. $\ket{\uparrow_L \uparrow_R}$ and $\ket{\downarrow_L \downarrow_R}$) for $\Delta_{\rm eff} < 0$. This is shown in Figure \ref{fig:magnetization}.

The edge field is now in the direction of anisotropy, so by the bosonization dictionary \eqref{eq:spin_field_correspondence}, it is a perturbation $\partial_x \varphi$ at the boundary. This is an exactly marginal perturbation, which can be absorbed by a singular gauge transformation leading to a line of RG fixed points with a non-universal critical exponents~\cite{0305-4470-31-12-003}. Phenomenologically, this gives the same result: we still have a ``superposition'' of boundary conditions, giving rise to double degneracy and edge magnetization. Now, however, the edge magnetization can have either algebraic or exponential decay, as shown in Figure \ref{fig:magnetization} (c), which is a feature of boundary fields in the $z$ direction for the XXZ spin chain. Both the $(Y)$- and $(Z)$-models are gapless SPTs with gapless edge modes. The primary difference is that the $(Y)$-model is more appealing from the field theory perspective, while the $(Z)$-model has more natural protecting symmetries (since in that case the $U(1)$ symmetry commutes with the SPT twisting unitary $U$).

\section{Conclusions}
We have demonstrated the existence of a gapless symmetry-protected topological phase, LL$^{\star}$. It has the crucial phenomenological feature of an SPT --- topological edge states --- so long as the protecting symmetries are preserved. Using standard field theory and numerical techniques, we have shown LL$^{\star}$ is a genuine phase, and thus robust to small perturbations. Using the language of boundary CFT, we have exactly solved the low-energy part of the model, which realizes an exotic ``superposition'' of conformal boundary conditions. These boundary conditions give rise to unusual edge effects in physical observables.

From a field theory point of view, gapped SPTs can be understood as non-linear sigma models with a bulk topological theta term with $\theta=2\pi$ \cite{PhysRevB.91.134404}. Gapless SPTs can then be obtained by tuning a subset of the protecting symmetries to criticality. It would be interesting to study how the kind of exotic surface criticality reported here arises in this language.

 There are multiple ways to interpret gSPTs. On the one hand, they provide a host of novel examples of surface criticality, where the anomalous edge properties produce effects similar to the so-called extraordinary transitions. The universality classes realized by these should be commonplace in the quantum phase transitions separating gapped SPTs and broken-symmetry phases. On the other hand, gSPTs can be thought of as new type of gapless topological matter. For example, twisting $U(1)$ gauge theories in 3D can lead to gapless topological phases in higher dimensions~\cite{PhysRevX.6.011034}. One could also partially gauge the symmetries~\cite{Levin2012}, which might produce gapless fractionalized matter. We therefore expect gSPTs to be a useful concept which paves the way to a complete understanding of gapless topological matter.

\begin{acknowledgements}
	We thank Ehud Altman and Ashvin Vishwanath for insightful discussions and Johannes Motruk for invaluable advice on DMRG. We acknowledge support from the Emergent Phenomena in Quantum Systems initiative of the Gordon and Betty Moore Foundation (T.S.) and NSF DMR-1507141 (D.P.).
\end{acknowledgements}

\bibliography{gapless_SPT}

\appendix

\section{Boundary Conformal Field Theory Partition Functions}

This Appendix will compute the partition functions of the free boson BCFT and derive the equation in Section \ref{sec:BCFT}. The computation is standard and relatively straightforward, but is detailed here for completeness. We will follow closely the references~\cite{oshikawa1997boundary, blumenhagen2009introduction}.

We would like to compute the partition function for the compactified free boson with various (conformally invariant) boundary conditions. Initially, we will work with Dirichlet boundary conditions, fixing $\varphi$ to $\varphi_a$ and $\varphi_b$ on the two ends. In order to follow more closely the CFT literature, we choose to work with a boson with compactification radius $r = \sqrt{g/2}$. Since time does not enter the Hamiltonian we may rescale $x \to x/v_F$ to eliminate the Fermi velocity and instead work on an interval of length $L/v_F$. With this convention, the Hamiltonian density is $\mathcal{H} = \frac{1}{2\pi}\left[ \left( \partial_x \theta \right)^2 + (\partial_x \varphi)^2 \right]$ and via the standard relation $\partial_x \theta = - \pi \Pi_\varphi = \partial_t \varphi$ \cite{giamarchi2003quantum}, this equivalent to the Lagrangian density $\mathcal{L} = \frac{1}{2\pi} \left( \partial_\mu \varphi \right)^2$.

The Dirichlet-Dirichlet partition function may now be computed on a cylinder with circumference $\beta$ and length $L/v_F$. It is convenient to use conformal invariance to reverse time and space: this is called the ``direct channel'' in the boundary CFT language. Define $\varphi = \varphi(x,t) \equiv \varphi(x+\beta,t)$ for $0 \le x \le \beta$ and $0 \le t \le L/v_F$. Since imaginary time is propagating along the original spatial direction, the partition function in this picture is not a trace, but is given by the amplitude
\begin{equation}
	Z^{DD}(\Delta \varphi) = \Braket{D(\varphi_a)|e^{- \frac{L}{v_F} H_\beta}|D(\varphi_b)},
	\label{eq:Z_DD}
\end{equation}
where $\ket{D(\varphi_{a,b})}$ are the conformal boundary states associated with Dirichlet boundary conditions, $\Delta \varphi = \varphi_a - \varphi_B$, and $H_\beta$ is the Hamiltonian for the free boson on the circle with circumference $\beta$. 

To proceed, expand the chiral components $\varphi(x,t) = \varphi_L(x^+) + \varphi_L(x^-)$ as
\begin{align*}
	\varphi_L(x^+) \ &=\ \frac{\hat{a_0}}{2} + \frac{\pi}{\beta}\left( r w + \frac{\hat{a^\dagger}}{2} \right)x^+\\
	\ &\quad +\ \frac{1}{2} \sum_{n=1}^\infty \left[ \frac{a_n}{\sqrt{n}} e^{-i n x^{+} \frac{2\pi}{\beta}} \right] + \frac{a_n^\dagger}{\sqrt{n}} e^{inx^+ \frac{2\pi}{\beta}},\\
	\varphi_R(x^-) \ &=\ \frac{\hat{a_0}}{2} + \frac{\pi}{\beta}\left( -r w + \frac{\hat{a^\dagger}}{2} \right)x^-\\
	\ &\quad +\ \frac{1}{2} \sum_{n=1}^\infty \left[ \frac{b_n}{\sqrt{n}} e^{-i n x^{-} \frac{2\pi}{\beta}} \right] + \frac{b_n^\dagger}{\sqrt{n}} e^{inx^- \frac{2\pi}{\beta}}.\\
\end{align*}
where $w$ is an integer winding number and $x^\pm =\pm x + t$. Since the constant mode $\hat{x}$ also obeys $\hat{x} \equiv \hat{x} +2 \pi \beta$, the conjugate momentum $p$ is quantized to an integer multiple of $1/r$. The new operators obey commutation relations $[a_n, a_m^\dagger] = [b_n, b_m^\dagger] = \delta_{nm}$ and $[\hat{x}, \hat{p}] = i$. We may then rewrite the Hamiltonian, in terms of the modes, either in the CFT picture as the generator of translations in the $t$ direction, or by putting the modes back into the Hamiltonian:
\begin{equation}
	H_\beta =\ \frac{2\pi}{\beta}\left[ (rw)^2 + \left( \frac{\hat{p}}{2} \right)^2 + \sum_{n=1}^\infty n a_n^\dagger a_n + n b_n^\dagger b_n - \frac{1}{12} \right].
	\label{eq:mode_Hamiltonian}
\end{equation}
The Dirichlet boundary states also take a simple form in terms of these modes~\cite{oshikawa1997boundary}
\begin{equation}
	\ket{D(\varphi_0)} = \frac{1}{\sqrt{2r}}\sum_{k \in \Z} e^{-ik \varphi_0/r} \exp\left[ -\sum_{n = 1}^{\infty} a_n^\dagger b_n^\dagger \right] \ket{(0,k)},
	\label{eq:Dirichlet_boundary_state}
\end{equation}
where $\ket{(w,k)}$ is the vacuum state with winding number $w$ and $\hat{p} = k/r$. For Neumann boundary conditions, by contrast, 
which can also be interpreted as Dirichlet boundary conditions for the dual field $\theta$, the boundary state is
\begin{equation}
	\ket{N(\varphi_0)} =\sqrt{r} \sum_{k \in \Z} e^{-i2rk \varphi_0} \exp\left[ +\sum_{n = 1}^{\infty} a_n^\dagger b_n^\dagger \right] \ket{(w,0)}.
	\label{eq:Neumann_boundary_state}
\end{equation}
These are related by the duality transformation $\varphi \leftrightarrow \theta$, $r \leftrightarrow 1/(2r)$ (or $g \leftrightarrow 1/g$ in the Luttinger parameter convention). 

We may now directly evaluate $\eqref{eq:Z_DD}$, and following~\cite{blumenhagen2009introduction}, we introduce the basis 
\begin{equation}
	\ket{\v{m}} = \ket{(m_1, m_2,\dots)} = \prod_{n=1}^\infty \frac{\left( a_n^\dagger \right)^{m_n}}{\sqrt{m_n!}} \ket{\Omega},
	\label{eq:clever_basis}
\end{equation}
where $\ket{\Omega}$ is the vacuum, and the factors were chosen so that $\braket{\v{n}|\v{m}} = \delta_{\v{n}\v{m}}$. Similarly, define $\ket{\widetilde{\v{m}}}$ with $a^\dagger_n \to b^\dagger_n$. It is also convenient to introduce the anti-unitary charge conjugation operator, $\mathcal{C}$ , whose action on the modes $\mathcal{C} a_n \mathcal{C}^{-1} = - a_n$, $\mathcal{C} b_n \mathcal{C}^{-1} = - b_n$ and performs complex conjugation $\mathcal{C} \, z\,  \mathcal{C}^{-1} = z^*$ for $z \in \C$. Using these states, we find
\begin{equation}
	\exp\left[ -\sum_{n=1}^\infty a_n^\dagger b_n^\dagger \right] \ket{\Omega} = \sum_{\{ \v{m} \}} \ket{\v{m}} \otimes \ket{ \mathcal{C}\, \widetilde{\v{m}}\,},
\label{eq:clever_basis_simplification}
\end{equation}
where $\{ \v{m} \}$ runs over all configurations of $m_n$. Notice that the state naturally separated into two chiral sectors.

Combining Equations \eqref{eq:Z_DD}, \eqref{eq:mode_Hamiltonian}, \eqref{eq:Dirichlet_boundary_state} and \eqref{eq:clever_basis_simplification} yields
\begin{align}
	Z^{DD}( \Delta \varphi) \ &=\ \frac{1}{2r} \sum_{k \in \Z, \{ \v{m}\}} e^{-ik \Delta \varphi/r} A \times B \times C,\\
	A \ &=\ \braket{k|\exp\left( -\frac{2\pi L}{v_F \beta}\left[ \left( rw \right)^2 + \left( \frac{\hat{p}}{2} \right)^2 - \frac{1}{12} \right] \right)|k}, \notag\\
	B \ &=\ \braket{\v{m}|\exp\left( -\frac{2\pi L}{v_F \beta}\left[ \sum_{n=1}^\infty n a_n^\dagger a_n \right] \right)|\v{m}},\notag\\
	C \ &=\ \braket{\mathcal{C} \,\widetilde{\v{m}}\,|\exp\left( -\frac{2\pi L}{v_F \beta}\left[ \sum_{n=1}^\infty n b_n^\dagger b_n \right] \right)|\mathcal{C}\,\widetilde{\v{m}}\,},\notag
\end{align}
where we have used the orthogonality property of the $\ket{k}$'s $\ket{\v{m}}$'s and $\ket{\,\widetilde{\v{m}}\,}$'s. We can now easily compute each of these terms. Since the winding number is always zero, $A = e^{-\frac{2\pi L}{v_F \beta}\left[ \left( \frac{k}{2r} \right)^2 - \frac{1}{12} \right]}$. Since $a_n^\dagger a_n$ is the number operator, we have $\exp\left( \sum_{n=1}^\infty n a_n^\dagger a_n \right) \ket{\v{m}} = \exp\left( \sum_{n=1}^\infty n m_n \right) \ket{\v{m}}$, which implies $B = e^{-\frac{2\pi L}{v_F \beta}\left[  \sum_{n=1}^\infty n m_n  \right]}$. Of course, the same logic applies to the other chiral sector, so $C = B$. The partition function is therefore
\begin{equation}
	Z^{DD}(\Delta \varphi) = \frac{1}{2r} \sum_{k, \{ \v{m} \}} e^{i k \Delta \varphi/r} \tilde{q}^{\frac{k^2}{8r^2}} \tilde{q}^{-1/24} \prod_{n=1}^\infty \tilde{q}^{n m_n},
\end{equation}
using the modular parameter $\tilde{q} = e^{2\pi i \tilde{\tau}} = e^{-\frac{4 \pi L}{v_F \beta}}$.
The $k$ and $\v{m}$ parts separate out. Focusing on the $\v{m}$ part,
\begin{equation}
	\sum_{\v{m}} \tilde{q}^{-\frac{1}{24}} \prod_{n=1}^\infty \tilde{q}^{n m_n} = \tilde{q}^{-\frac{1}{24}} \prod_{n=1}^\infty \sum_{m_n = 0}^\infty \tilde{q}^{nm_n} = \frac{1}{\eta(\tilde{q})}. 
\end{equation}
where $\eta(q) = q^{\frac{1}{24}} \prod_{n=1}^\infty (1- q^n)$ is the Dedekind $\eta$-function. The $k$-part is naturally expressed in terms of the Jacobi $\theta$-function~\cite{elias2003complex}
\begin{equation}
	\Theta(z | \tau) = \sum_{k \in \Z} q^{\frac{1}{2} k^2} e^{2\pi i k z}
= \sum_{k \in \Z} \left( e^{2\pi i \tau }\right)^{\frac{1}{2} k^2} e^{2\pi i k z}.
	\label{eq:Jacobi_theta_function}
\end{equation}
Hence the partition function is
\begin{equation}
	Z^{DD}(\Delta \varphi) = \frac{1}{2r} \frac{1}{\eta(\tilde{q})} \Theta\left(\frac{\Delta \varphi}{2\pi r} \; \Bigg| \; -\frac{1}{2r^2} \frac{L}{v_F \beta} \right).
	\label{eq:Z_DD_II}
\end{equation}

Finally, we use the modular properties of the $\eta$ and $\Theta$ functions to rewrite the partition function as a sum of Boltzmann weights to read off the operator content of the theory. For $\operatorname{Im} \tau > 0$~\cite{elias2003complex} 
\begin{align*}
	\Theta(z | -1/\tau) \ &=\ \sqrt{\tau/i} \; e^{i \pi \tau z^2} \Theta\left( z \tau | \tau \right),\\
	\eta(-1/\tau) \ &=\ \sqrt{\tau/i} \eta(\tau).
\end{align*}
Applying these and simplifying yields
\begin{equation}
	Z^{DD}(\Delta \varphi) = \frac{1}{\eta(q)} \sum_{k \in \Z} q^{g \left( k+z \right)^2},
	\label{eq:Z_DD_III}
\end{equation}
with conjugate modular parameter $q = e^{-\beta \frac{\pi v_F}{L}}$, $z = \Delta \varphi/(2\pi r)$, and we have substituted back in the Luttinger parameter $g = 2r^2$. This is precisely Equation \eqref{eq:Z_gTrivial}.

In the CFT language, the state-operator correspondence tells us each term in the sum \eqref{eq:Z_DD_III} represents one operator, whose scaling dimension is~$g (k+z)^2$. Notice that if we increase $z \to z + 1$, the partition function is unchanged, which comports with the fact that $\varphi$ is compactified. The role of $\eta$ is to generate descendant fields. It is the generating function for partitions numbers $\prod_{n=1}^\infty \frac{1}{1-q^n} = \sum_{n=1}^\infty p(n) q^n$~\cite{elias2003complex}. 

Expanding the partition function as
\begin{equation}
	Z^{DD}(\Delta \varphi) = \sum_{k \in \Z, n\ge 1} p(n) q^{-\frac{1}{24} + g(k+z)^2 + n},
	\label{eq:Z_DD_IV}
\end{equation}
and interpreting these as Boltzmann weights, we find the boundary spectrum of the theory is
\begin{equation}
	E_{k,n} = \frac{\pi v_F}{L} \left[ -\frac{1}{24} + g(k+z)^2 + n \right],
	\label{eq:spectrum}
\end{equation}
with multiplicity $p(n)$.

Now that we have done the Dirichlet case, the case with Neumann boundary conditions at both ends is simple. We simply substitute $\Delta \varphi \to \Delta \theta$ and $g \to 1/g$ in Equation \eqref{eq:Z_DD_IV} to find
\begin{equation} 
	Z^{NN}(\Delta \theta) = \frac{1}{\eta(q)} \sum_{k \in \Z} q^{\left( k+ w \right)^2/g},
	\label{eq:Z_NN_I}
\end{equation}
where $w = \Delta \theta r/ \pi$. When we pin a spin at the edge to point up (down), that is equivalent to pinning $\theta = 0$ ($\theta = \pi/2r$). The superposition of up and down at both ends is therefore
\begin{equation}
	Z_\text{gSPT} = Z^{NN}(0) + Z^{NN}(0) + Z^{NN}(\pi/2r) + Z^{NN}(-\pi/2r),
\end{equation}
so that
\begin{equation}
	Z_\text{gSPT} = \frac{2}{\eta(q)} \sum_{m \in \Z} \left( q^{m^2/g} + q^{\left( m+\frac{1}{2} \right)^2/g} \right).
\end{equation}
This is precisely Eq.~\eqref{eq:Z_gSPT_II}.

\end{document}